\documentclass[pre,showpacs,twocolumn]{revtex4}
\usepackage{epsfig}
\font\tams                   = cmmib10
\font\kleinhalbcurs          = cmmib10 scaled 833
\font\bxf                    = cmbx10
\font\sevenbf                = cmbx7
\def\vec#1{{\textfont1=\tams\scriptfont1=\kleinhalbcurs
\textfont0=\bxf\scriptfont0=\sevenbf
\mathchoice {\hbox{$\displaystyle#1$}}{\hbox{$\textstyle#1$}}
{\hbox{$\scriptstyle#1$}}{\hbox{$\scriptscriptstyle#1$}}}}

\def \v0{\vec{0}}

\def \re{\mbox{\rm e}}
\def \Prob{\mbox{\rm Prob}}
\def \VaR{\mbox{\rm VaR}}

\def \be{\begin{equation}}
\def \ee{\end{equation}}
\def \bea{\begin{eqnarray}}
\def \eea{\end{eqnarray}}
\def \nn{\nonumber}

\begin{document}
\pagestyle{empty}
\title{\Large\bf Functional Correlation Approach to Operational Risk
in Banking Organizations\footnote{The views presented in this
paper are those of the authors and do not necessarily represent
models or policies  of Dresdner Bank AG.}}

\author{Reimer K\"uhn$^1$ and Peter Neu$^2$}
\affiliation{$^1$Institut f\"ur Theoretische Physik, Universit\"at Heidelberg,
Philosophenweg 19, D-69120 Heidelberg, Germany\\
$^2$Group Risk Control, Dresdner Bank AG,  J\"urgen-Ponto-Platz
1, D-60301, Frankfurt, Germany }

\date{\today}

\begin{abstract}
A Value-at-Risk based model is proposed to compute the adequate
equity capital necessary to cover potential losses due to
operational risks, such as human and system process failures, in
banking organizations. Exploring the analogy to a lattice gas
model from physics, correlations between sequential failures are
modeled by as functionally defined, heterogeneous couplings
between mutually supportive processes. In contrast to traditional
risk models for market and credit risk, where correlations are
described by the covariance of Gaussian processes, the dynamics
of the model shows collective phenomena such as bursts and
avalanches of process failures.
\end{abstract}

\maketitle

\pagestyle{plain}

\section{Introduction}
Risk management has become increasingly important in financial
institutions over the last decade. Since the publication of JP
Morgan's RiskMetrics$^{\rm{TM}}$~\cite{RM} in the nineties, Risk
Management and Risk Control departments in banks have grown
significantly in size and importance. The task is to fulfill
regulatory requirements, to add transparency about a bank's risk
profile by a quantitative assessment of risks, to develop the
necessary IT-solutions which allow to process the huge amount of
data of a bank, and, finally, to integrate this information in a
risk-return (RoRAC = Return on Risk-Adjusted Capital) based
steering process of the bank. Ultimately, a proper risk
management and risk control process is recognized by rating
agencies and investors so that shareholder value is added to the
bank.

Banks first focused on controlling potential losses due to market
fluctuation, such as changes in the S\&P 500 stock index,
changes in interest and currency exchange rates, which is termed
market risk. Internal market risk models are nowadays rather
matured and accepted by regulators for the calculation of the
required capital to be held as buffer against such losses. In
contrast to these elaborated statistical models for market risks,
credit risks (i.e., risks due to defaulted obligors) have to be
covered  by simply 8\% capital of the bank's risk-weighted assets.
Implicitly, this charge also includes other risks such as
operational risks. Since the New Basel Accord on Capital Adequacy
issued by the Basel Committee on Banking Supervision in February
and September 2001 \cite{BISIIa, BISIIb, BISIIc}, known as Basel
II, it is clear that regulators will demand banks to hold equity
capital against operational risks explicitly.

A common industry definition of operational risk (OR) is the risk
of direct or indirect losses resulting from inadequate or failed
internal processes, people and systems or from external events
\cite{Frachot01}. See \cite{Vandenbrink} for a practice-oriented
introduction to the issue. Possible OR-risk categories are
\cite{Vandenbrink}:
 (i) human processing errors, e.g., mishandling of software
applications, reports containing incomplete information, or payments
made to incorrect parties without recovery,  (ii) human decision
errors, e.g., unnecessary rejection of a profitable trade or wrong
trading strategy due to incomplete information,  (iii) (software
or hardware) system errors, e.g., data delivery or data import is not
executed and the software system performs calculations and generates
 reports based on incomplete data,  (iv) process design error,
e.g., workflows with ambiguously defined process steps,  (v)
fraud and theft, e.g., unauthorized actions or credit card fraud, and
(vi) external damages, e.g., fire or earthquake.

Thinking of theses categories as ``operational risk processes" it
is clear that there are {\em functionally defined dependencies}
between individual processes, which all together bring a big
organization to work. Consider the following example for
illustration: a system error leads to an incomplete data import
into a risk calculation engine, resulting in a wrong calculation
of risk figures, and eventually to a human decision error by the
trader, who closes a possibly profitable position unnecessarily to
reduce a risk which in fact does not exist.

In the end misleading or lagging information, or system and
workflow failures will always result in financial loss for a
bank. Indeed, practitioners have recognized these dependencies in
operational risk events and mandated units like the internal audit
and risk control departments to control processes for the bank,
and generated functions like a Chief Operating Officer (COO) to
optimize them. Operational risk error trees between the above
categories have been formalized in \cite{Vandenbrink} in more
detail.

Since the mid nineties financial markets have also attracted
physicists in academia. One of the main reason is that financial
time series exhibit several statistical peculiarities, many of
them being common to a wide variety of different markets and
instruments. As such they could possibly be ``universal", i.e.,
independent of market details like instruments, country, and
currency, and be the signature of collective phenomena in
financial markets (see \cite{Bouch0,Bouch1,Bouch2, Sornette1,
Stanly1} and references therein). Collective phenomena have
been widely studied in physics in the context of phase
transitions. Collective phenomena are often responsible for
insensitivity of overall system behavior to details of an
underlying dynamics. Specifically at phase transitions they
give rise to power-law behavior, scale-invariance and
self-similarity. Similar properties of financial time series
might therefore well be understood as a consequence of agents
in a market acting collectively.

Bringing together ideas from physics about collective phenomena
and best industry practice for risk measurement the present paper
details a possible statistical approach to determine the necessary
equity capital to be held by banks to cover losses due to
operational risks. In physical terms our model resembles a
lattice gas with heterogeneous, functionally defined couplings.
In such a description, bursts and avalanches of process failures
correspond to droplet formation associated with a first order
phase transition.

The paper is organized as follows. In Sect. \ref{VaR_Concept} we
describe the Value-at-Risk concept for risk management and
control. In Sect. \ref{OR} we describe the approaches discussed
in the context of Basel II  for operational risk measurement. A
new approach based on functionally dependent correlations giving
rise to collective behavior is introduced in Sect.
\ref{Sect_LatticeGas}. Finally, Sect. \ref{Sect_Conclude}
summarizes our results.

\section{The Value-at-Risk Concept} \label{VaR_Concept}

Risk management in banks is based on diversification, hedging and
equity capital as loss buffer. The bank charges its customers a
premium for its risks so that (expected) losses in one market
segment are on average compensated by profits in others. Other
risks, especially market and increasingly credit risks, are hedged
(insured) via the derivative market. Unexpected losses, which are
not diversified or hedged,  are covered by the bank's equity
capital. How much capital a bank needs to cover its risks is
determined by the so-called ``Value-at-Risk" (VaR). VaR can be
defined as the worst loss in excess of the expected loss that can
happen under normal market conditions over a specified horizon
$T$ at a specified confidence level $q$. More formally, VaR
measures the shortfall from the $q$-quantile of the loss
distribution in excess of the expected loss, $EL$, within the
time period $T$ discounted at the risk-free rate $r$ to time $t =
0$:
\be \VaR_{q,T} = \left(Q_{q}[L(T)] - EL \right)\re^{-rT}\ ,
\label{I3}\ee
where the $q$-quantile worst case loss, $Q_{q}[L(T)]$, is defined
at confidence level $q$ through
 \be \Prob\left(L(T) > Q_{q}[L(T)]\right)= 1-q\ . \label{I1}\ee

As indicated in (\ref{I3}), VaR depends on the confidence level
$q$ and the risk horizon $T$. The choice of these parameters
depends on the application. If VaR is simply used to report or
compare risks, these parameter can be arbitrarily chosen, as long
they are consistent. If, however, VaR is used as a basis for
setting the amount of equity capital, the parameters must be
chosen with extreme care: the confidence level must reflect the
default probability of the bank within the risk horizon, and the
risk horizon must be related to the liquidation period of risky
assets, recovery time of ill-functioning processes, or,
alternatively, to the time period necessary to raise additional
funds. This explains why regulators have chosen a high confidence
level of 99\% and a 10-day horizon to determine the minimum
capital level for market risks. For credit risks and capital
allocation, banks choose $q$ and $T$ even higher about  99.95\%
and one year, respectively.

In the financial industry there exist established statistical
models for market and credit risk. Statistical models for
operational risk start now to be discussed in the risk management
community, especially in the context of Basel II \cite{BISIIb}.
Whereas internal market risk models are already recognized by
regulators and are also used in banks for capital allocation,
regulators are much more critical about internal statistical
models for credit and operational risk. This is clearly less
related to the mathematical complexity --- although credit and
operational risks are more difficult to model than market risks
--- but to problems with respect to input data which are much
harder to validate than in the case of market risk.

\section{Industry Standards for Operational
Risk Measurement}\label{OR}

The
Basel Committee for Banking Supervision has proposed three
alternative approaches to  operational risk measurement 
\cite{BISIIc}: The ``Basic-Indicator Approach (BIA)",
 the
``Standardized Approach (SA)", and the ``Advanced Measurement
Approach (AMA)". In the BIA  the required capital for operational
risk is determined by multiplying a single financial indicator,
which is gross income (interest, provision, trading, and other
income) by a fixed percentage (called the $\alpha$-factor). The
SA differs from the latter in that banks are allowed to choose
business line specific weight factors, $\beta_k$, for the gross
income indicator, $I_k$, of the $k^{\rm th}$ business line. The
total regulatory capital charge, $RC$, is the simple sum of the
capital required per business line,
 \be RC = \sum_k \beta_k \times I_k\ .\ee
The weight factors $\alpha$ and $\beta_k$ are calibrated such
that the
required regulatory capital for operational risk would be 
17 -
20\% of the current regulatory capital on bank average standards.

The AMA consist of three sub-categories: The ``Internal
Measurement Approach (IMA)", the ``Loss Distribution Approach
(LDA)" and the ``Scorecard Approach (SCA)". It is a more advanced
approach as it allows banks to use external and internal loss data
as well as internal expertise.

In the IMA the required capital is calculated as the sum over
multiples of the expected loss per OR-risk category/business line
cell
 \be RC = \sum_{i,k}
 \gamma_{ik} \times EL_{ik}\ ,\ee
where $i$ is the risk category and $k$ the business line. The
expected loss is quantified as the product of the annual OR-event
probability, an exposure indicator per business line and risk
category, and the loss percentage per exposure. All parameter
estimates have to be disclosed to the supervisors. Since the
$\gamma$-factor is computed on an industry based distribution, it
will be possible to adjust the capital charge by a  risk profile
index, which accounts for the bank's specific risk profile
compared to industry.

The LDA and SCA are very similar as both approaches are based on a
statistical VaR-model. Details of the LDA approach are outlined in
ref.~\cite{Frachot01}. In both approaches the bank estimates for
each risk category/business line cell the probability
distributions of the annual event frequency and the loss severity
(= exposure $\times$ loss fraction per exposure). The difference
between the LDA and the SCA is that in the former only internal
or external historical loss data are used for estimating the
distribution functions. In addition to this, banks are also
allowed to apply expert knowledge to estimate the distribution
functions in the SCA. This is a forward looking approach. It is
particularly suited for operational risk, as processes that have 
failed are
usually changed; hence historical loss data could provide
potentially misleading information. Even if banks have an exhaustive
internal database of losses, it can hardly be considered as
representative of extreme losses. Hence, expert assessments and
external loss databases are necessary. The problem with the
latter data source is that external historical losses must be
scaled to fit the balance sheet of the bank (it must be possible
that the losses can occur in the bank).

Popular choices for the loss severity distribution functions are
the lognormal, Gamma, Beta, Weibull distribution. Common choices
for the loss frequency distribution function are the Poisson or
negative binomial distribution. In a top-down approach different
OR-risk categories are assumed to be independent. For each
business unit and for each OR-category the OR-loss is simulated
in a Monte-Carlo simulation by drawing a realization $N_{ik}$
from the loss frequency and sampling $N_{ik}$ realizations of the
loss severity $X_{ik}^{m}$ ($m = 1, \ldots, N_{ik}$). The loss in
such a sample is
\be \label{OR1} L_{ik} = \sum_{m=1}^{N_{ik}} X_{ik}^{m}\ . \ee
Drawing a histogram of outcomes of $L_{ik}$ provides the loss
distribution function per risk category/business line cell. The
Value-at-Risk is read off from the tail in excess to the expected
loss as described in Sect. \ref{VaR_Concept}. Due to the
assumption of statistical independence, the loss distribution can
also be calculated analytically as the convolution product of the
loss frequency and the loss severity distribution. The required
capital for the bank as a whole can either be calculated as the
simple summation of the capital charges across each of the risk
category/business line cell. This is the method given by the
Basel Committee on Banking Supervision in the Internal
Measurement Approach. Or, the MC-sampling can be extended beyond
the risk category/business line cell by $L = \sum_{i,k} L_{ik}$,
which takes diversification between the risk category/business
line cell  into account.

A critical point which concerns all presently discussed
approaches is the correlation between OR-losses. In this paper
our focus will be how correlations and dependencies between
OR-risk events can be integrated in the LDA/SCA.

\section{Functional Correlation Approach for Operational Risk}
\label{Sect_LatticeGas}

Since Markowitz's centennial work on portfolio theory
\cite{Markowitz52,Markowitz59}, {\em diversification} and {\em
dependencies} between risk events are modeled by the covariance
of stochastic processes. Because empirically only the mean and
the covariance of these processes are reliably determined from
market data, it is common practice to choose correlated Gaussian
white noise for modeling correlations. As a consequence the loss
distribution is unimodal with frequent small losses and a few
extreme losses, which --- dependent on the distribution of the
loss severity --- are responsible for fat tails in the loss
distribution. Collective losses or even crashes such as burst and
avalanches of losses are not contained in this description.

A main point in this paper is that this stochastic dependency of
risk events is not sufficient for all risk categories: one
frequently also observes {\em direct, functional and
non-stochastic dependencies}. Functional dependency between risk
events is most pronounced in operational risk events. Processes
in a (large) organization are usually organized so as to mutually
support each other. Thus, if a process fails, this will usually
be detrimental to other processes relying on receiving input or
support of some sort from the failing process in question, so
that they run a higher risk of failing as well. It  therefore
seems inadequate to model operational risk events individually per
risk category/business line cell and aggregate losses afterwards
over some covariance matrix, which would be the choice when
approaching operational risks analogously to market risks. In the
following we extend the LDA/SCA by taking the
{\em functional
dependencies\/} between processes into account.

We consider a simple two state model here, i.e., a processes can
be either up and running or down. For the process corresponding
to the OR-event $i$ we designate these states as $n_{i}=0$ and
$n_{i}=1$, respectively. In following we will skip the business
line index $k$ for simplicity.

The interest is in obtaining reliable estimates of the statistics
of processes that are down at any time and of the statistics of
losses incurred at any time. As the loss severity incurred by a
given process going into the down state may vary randomly from
event to event, solving the latter problem requires convolving the
statistics of down-events with the loss severity distribution
related to the process failures.

The reliability of individual processes will vary (randomly)
across the set of processes, and so will the degree of functional
interdependence. These random heterogeneities constitute an
element of quenched disorder, whereas the loss severities incurred
by down processes constitute an element of annealed disorder as
they are (randomly) determined anew from their distribution each
time a process goes down. An appealing feature for the modeler of
operational risks therefore is the {\em independence\/} of the
dynamic model of the interacting processes and the loss severity
model (i.e. the estimate of the PDFs of loss severity incurred by
individual process failures.) A typical assumption for the latter
is to take them as being distributed according to a log-normal
distribution with suitable parameters for means and variances,
which we will choose in the following.

\subsection{Dynamics}

To motivate the dynamics of the functional approach, note that all
processes need a certain amount of ``fueling" or support in order
to maintain a functioning state for the time increment $t \to t +
\Delta t$ within the risk horizon, $t \in [0,T)$ (think of human
resources,  information, input from other processes, etc.). Here,
only the generic features of the model shall be outlined. Hence,
the increment $\Delta t$ is chosen such that all processes can
fully recover within this time interval, i.e., the state $n_i$ of
each process can flip each time step. For practical applications
in banks, one would model the recovery process more carefully:
specific death-period after the failure of the $i^{\rm th}$
process would be considered,  and one would differentiate between
process failures being discovered and adjusted up to a certain
cut-off time, e.g., end-of-day, at which a process would have
been completed \cite{Vandenbrink}. These features are not generic
and can only be discussed related a specific OR-event under
consideration.

We denote by $h_i(t)$ the total support received by process $i$
at time $t$, and choose it to take the form
\be h_i(t) = \vartheta_i - \sum_j w_{ij} n_j(t) + \eta_i(t)\ .
\label{hioft} \ee
That is, it is composed of (i) the {\em average\/} total support
$\vartheta_i$ that would be provided by a fully operational
network of processes (in which $n_i(t)=0$ for all $i$). This
quantity is (ii) {\em diminished\/} by support that is missing
because of failing  processes which normally feed into the process
in question; (iii) lastly, there are fluctuations about the
average which we take to be correlated Gaussian white noise with
--- by proper renormalizing $\vartheta_i$ and $w_{ij}$ --- zero
mean and unit variance. Correlated Gaussian noise is introduced to
model equal-time cross correlations between OR-risk categories in
analogy to the approach proposed by the Basel Committee for
Banking Supervision for credit risk,
\be \eta_i(t) = \sqrt{\rho}\, Y(t) +
\sqrt{1-\rho}\,\epsilon_i(t)\ , \label{etai}\ee
where $Y(t)\sim {\cal N}(0,1)$ is a common factor for all OR-risk
categories with equal-time correlation coefficient $\rho$, and the
$\epsilon_i(t)\sim {\cal N}(0,1)$ are idiosyncratic terms. 

Note that non-linear effects could be included by modifying
(\ref{hioft}) to $h_i(t) = \vartheta_i - \sum_j w_{ij} n_j(t) -
\sum_{j,k} w_{ijk} n_j(t)n_k(t) - \dots + \eta_i(t)$. Note also that,
as in credit risk modeling, the common factor $\sqrt{\rho}\, Y(t)$
could further be decomposed into sector-contributions $\sqrt{\rho}\,Y(t)
\to \sum_k \beta_{ik} Y_k(t)$ so as to describe more complicated 
equal-time correlations. To keep this treatment transparent, we will 
present
the formalism without these extensions.

Process $i$ will fail in the next time instant $t+\Delta t$, if
the total support for it falls below a critical threshold. By
properly renormalizing $\vartheta_i$, we can choose this
threshold to be zero, thus ($\Theta$ is the step-function:
$\Theta(x) = 1$ for $x \ge 0$ and 0 else)
\bea n_i(t+\Delta t) &=& \Theta\Big(-  \vartheta_i + \sum_j w_{ij}
n_j(t) \nn\\& &- \sqrt{\rho}\, Y(t) -\sqrt{1-\rho}\,\epsilon_i(t)\Big)
\ . \label{nioft+} \eea
The losses incurred by process $i$ are then updated according to
\be L_i(t+\Delta t) = L_i(t) + n_i(t+\Delta t) X^i_{t+\Delta t}\ ,
\ee
where $X^i_{t+\Delta t}$ is randomly sampled from the loss
severity distribution for process $i$. Note that the {\em process
dynamics\/} is {\em independent\/} of assumptions concerning
their loss severity distributions within the present model.

One can integrate over the distribution of {\em idiosyncratic\/}
noises to obtain the conditional probability for failure of
process $i$ given a configuration $n(t) = \{n_i(t)\}$ of
down-processes  and a realization of the common factor $Y(t)$ at
time $t$,
\bea \label{pdown}  \langle n_i(t+\Delta t)\rangle_{n(t), Y(t)}
&& \nn\\ && \hspace{-3cm}\equiv
\Prob\left.\Big(n_i(t+\Delta t)=1 \right|n(t),
Y(t)\Big)\\
&& \hspace{-3.cm}= \Phi\left(\frac{\Phi^{-1}(p_{i}) + \sum_{j}w_{ij}\,
n_j(t) - \sqrt{\rho}\,Y(t) }{\sqrt{1-\rho}} \right)\nn\ .\eea
Here $\Phi(x)$ denotes the cumulative normal distribution. 
Note that we have set $\vartheta_i = - \Phi^{-1}(p_{i})$, where
$p_{i,\Delta t}\equiv p_i$ is the {\em unconditional\/} expected
probability for process failures within the time-increment
$\Delta t$. This is consistently justified by setting  $n_j(t) =
0$ for all $j$ and $\rho = 0$ in Eq. (\ref{pdown}). Note that up
to the functional term ``$\sum_{j}w_{ij}\, n_j(t)$" this approach
corresponds to the approach adapted by the Basel Committee for
Banking Supervision for credit risk \cite{BISIIc}.

The couplings $w_{ij}$ can be determined by considering the
transition probabilities, $p_{ij,\Delta t}\equiv p_{ij}$, for
process $i$ failure within the time-increments $\Delta t$, given
that in the configuration at time $t$ process $j$ is down,
whereas
all other processes are running, and $Y(t) = 0$. Introducing the
shorthand ${\cal C}_j$ for this configuration, we can write
 
\bea
p_{ij} &=& \Prob\left.\Big(n_i(t+\Delta t)=1 \right|{\cal
C}_j\Big)\nonumber\\ &=& \Phi\Big(\frac{\Phi^{-1}(p_{i}) +
w_{ij}}{\sqrt{1-\rho}}\Big)\ . \label{pij} \eea
This leads to
 \be
w_{ij} = \sqrt{1-\rho}\,\Phi^{-1}(p_{ij}) - \Phi^{-1}(p_{i})
 \ .
\label{fixwij} \ee
Analogous identities would be available for determining higher
 order
connections $w_{ijk}$, if nonlinear effects were taken into
 account. Note
that the probabilities for process failure depend
 only on the increment
$\Delta t$ and not on the time $t$ due to
 the stationarity of the dynamics.
 
To illustrate how these parameter are fixed in practice, consider
the following. Either from a historical loss database, or from an
expert assessment the following two questions must be answered per
OR-risk category and business line:

\begin{enumerate}
\item What is the expected period, $\langle \tau_i \rangle$, until process
$i$ fails for the
 first time in a fully operative environment, and
\item given that only process $j$ has failed, what is the expected
period, $\langle\tau_{ij}\rangle$, for process $i$ to fail also?
\end{enumerate}

Noting that with $\Prob({\rm failure\ at}\ z\Delta t) =
\big(1-p_{i}\big)^{z-1}p_{i}$ one finds that
\be \langle \tau_i \rangle = \sum_{z = 1}^\infty z\Delta t \,
\big(1-p_{i} \big)^{z-1}p_{i} = \frac{\Delta t}{p_{i}} \ , \ee
and analogously,
\be \langle\tau_{ij}\rangle = \sum_{z = 1}^\infty z\Delta t\,
\big(1-p_{ij} \big)^{z-1}p_{ij} = \frac{\Delta t}{p_{ij}} \ . \ee
These identities express the $p_{i}$ and $p_{ij}$ in terms of
estimated average times of failure, and are used to fix the model
parameters completely. Note that according to (\ref{pij}) $p_{ij}$
can  be interpreted as a non-equal time correlation for process
failures.

Note also that, incidentally, the dynamics (\ref{nioft+})
resembles that of a lattice gas (defined on a graph rather than
on a lattice), the $n_i$ denoting occupancy of a vertex, the
$w_{ij}$ interactions, and $\vartheta_i$ taking the role of
chemical potentials regulating a-priori occupancy of individual
vertices. The present system is heterogeneous in that (i) the
$\vartheta_i$ vary from site to site, (ii) the couplings $w_{ij}$
have a functional rather than regular geometric dependence on the
indices $i$ and $j$ designating the vertices of the graph.
Moreover, in the physics context, one usually assumes noise
sources other than Gaussian so that cumulative probabilities are
described by Fermi-functions rather than cumulative normal
distributions as above. The quantitative difference is minute,
however.

The model dynamics as such cannot be solved analytically for a
general heterogeneous network. We shall resort to Monte-Carlo
simulations to study its salient properties. The main qualitative
features can, however, also be observed  in the simplified
situation of a homogeneous network consisting of identical
processes, having the same connectivity at each node.  A
mean-field analysis of such a simplified situation will be given
as well. As the presence of the common factor expressed by the
$\rho$-term in Eq. (\ref{etai}) would influence only quantitative
details of the system's behavior, we will further present the
analysis without correlation to the common factor by setting
$\rho \equiv 0$.

\subsection{Key Features}

Key features of the collective behavior of networks of
interacting processes can easily be anticipated either directly
from a discussion of the dynamic rules, or from  the analogy with
the physics of lattice gasses.

Consider a network in which the unconditional probabilities for
process failures, $p_{i}$, are small, but process interdependence
is large and consequently {\em conditional\/} probabilities for
process failures, $p_{ij}$, are sizeable. In such a situation,
spontaneous failure of individual processes  may induce
subsequent failures of other processes with sufficiently high
probability so as to trigger a breakdown of the whole network.
If, on the other hand, process interdependence remains below a
critical threshold value, individual spontaneous failures will
not have such drastic consequences, and the whole network will
remain in a stable overall functioning state.

Of particular interest for the risk manager is the case, in which
process interdependence is low enough to make a {\em
self-generated break-down\/} of the network extremely unlikely,
but parameters are nevertheless such that a stable overall
functioning state of the network {\em coexists\/} with a phase in
which nearly the complete network is in the down state (two phase
coexistence). In such a situation, it may be {\em external
strain\/} which can induce a transition from a stably functional
situation to overall breakdown. Analogous mechanisms are believed
to be responsible for occasional catastrophic breakdowns in
bistable ecosystems \cite{Nature01}.

With increasing unconditional probabilities for process failures it
becomes meaningless to distinguish between an overall functioning, and a
non-functioning phase of the network, two--phase coexistence ceases to exist
--- as in (lattice) gasses --- at a critical point.

\subsection{Simulations}

In the following we validate some of our intuitions about global
network behavior using Monte-Carlo simulations.

The Monte-Carlo dynamics can either be conceived as parallel
dynamics (all $n_i$ are at each time step simultaneously updated
according to (\ref{nioft+}) or (\ref{pdown})), or as (random)
sequential dynamics  (only a single $n_i$ is (randomly) selected
for update according to (\ref{nioft+}) in any given time-step, in
which case the time increment must scale with the number $N$ of
processes in the net as $\Delta t \sim N^{-1}$).

For the analysis of operational risks, losses are accumulated
during a Monte-Carlo simulation of the process dynamics over the
risk horizon, $T$. Runs over {\em many\/} risk horizons then allow
to measure loss distribution functions for individual processes
within the network of interacting processes, or of business units
or the full network by appropriate summations.

For the simulations, we choose a random setting, i.e.,
unconditional failure probabilities are taken to be homogeneously
distributed in the interval $[0,p^{\rm max}]$ and we determine
random conditional failure probabilities as $p_{ij} =
p_{i}(1+\varepsilon_{ij})$, with $\varepsilon_{ij}$ homogeneously
distributed in $[0, \varepsilon^{\rm max}]$, which fixes the ratio
$(p_{ij}/p_{i})^{\rm max}$.

Fig. 1 shows a situation where a functional network coexists with
a situation in which the network is completely down, and
parameters are such that spontaneous transitions between the
phases are not observed during a simulation.
The upper track shows the
loss record of a system initialized in the ``all down" state whereas
the lower track exhibits the loss record of the {\em same\/} network
initialized in the fully functioning state.

The loss distribution for the {\em functional\/} network is unimodal with a
bulk of small losses and a fat tail of extreme losses, which are driven by
the loss severity  distribution.

By increasing the functional interdependence at unaltered unconditional
failure probabilities, the functioning state of the network becomes unstable.
A spontaneous transition into the `down' state is observed during a single run
of 50000 Monte-Carlo steps. Two interesting features about this transition
to complete breakdown deserve mention:  (i) the time to breakdown can
vary within very wide limits (we have not attempted to measure the distribution
of times to breakdown and its evolution with system parameters such as range
of values for conditional failure probabilities),  (ii) there are no
detectable precursors to the transition; it occurs due to large spontaneous
fluctuations carrying the system over a barrier, in analogy to droplet
formation associated with first order phase transitions.

We should like to emphasize that realistically the system dynamics
after an overall break-down of a process network would no longer be
the spontaneous internal network dynamics: recovery efforts would
be started, increasing support for each process by a sufficient
amount such as to reinitialize the network in working order.

Repeated spontaneous transitions in both directions can be observed only in a
rather small network (Fig. 2). The  corresponding loss distribution will be
bimodal.

\begin{figure}[p]
{\centering
\epsfig{file=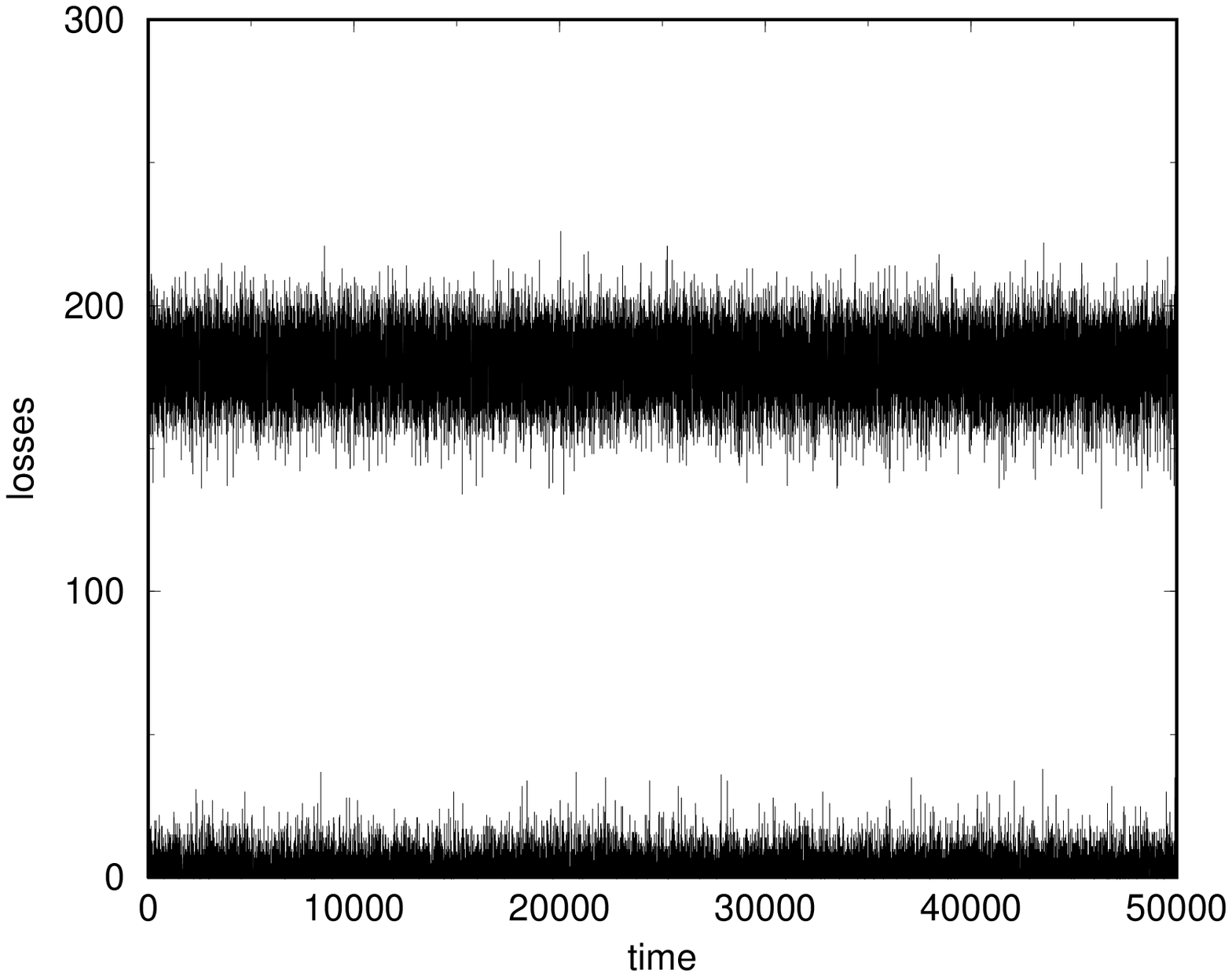,width=6.5cm} \hfil
\epsfig{file=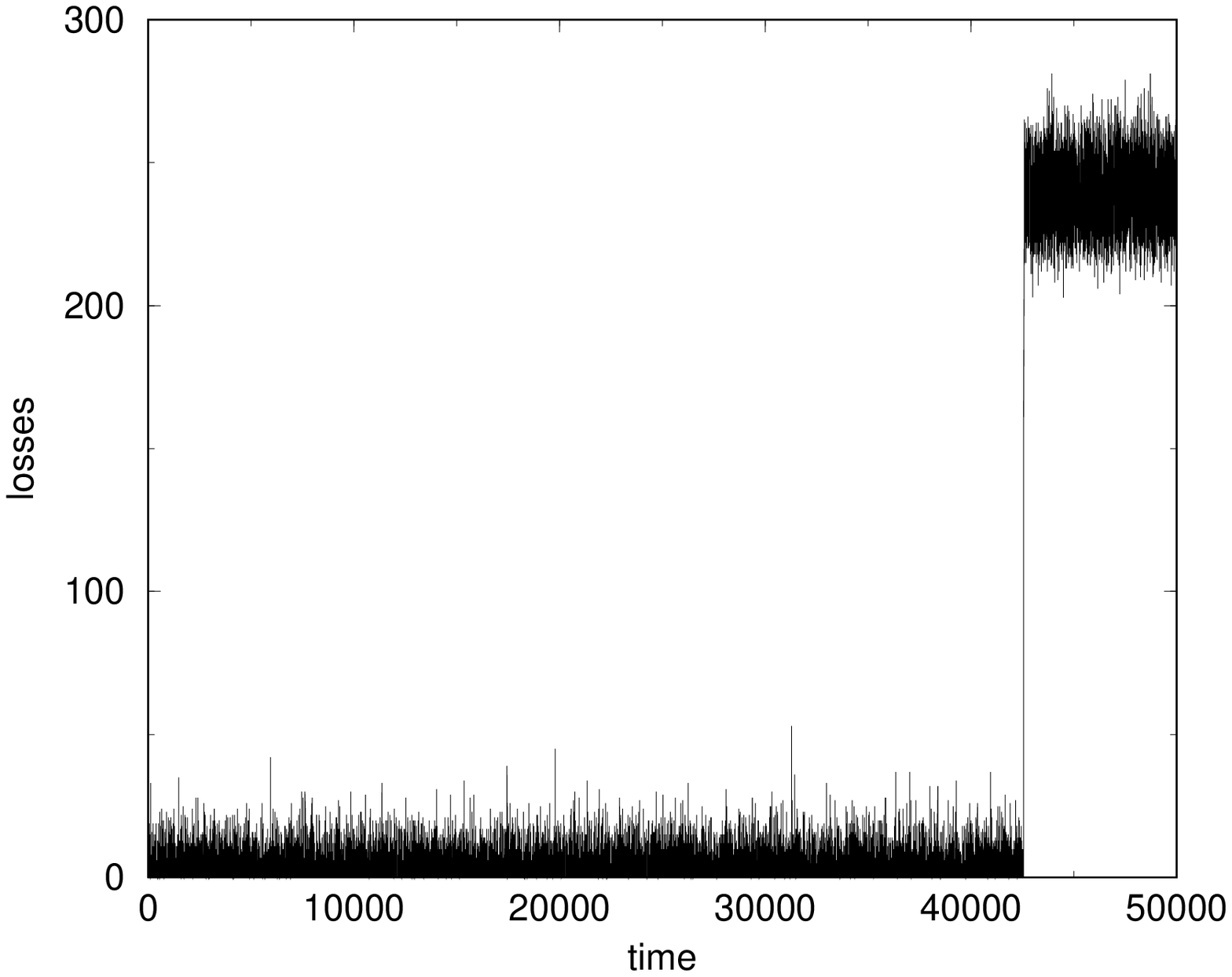,width=6.5cm}
\par}
\caption[]{Loss record for a system of $N=50$ interacting processes with
(first panel) $p^{\rm max}=0.02$ and $(p_{ij}/p_{i} )^{\rm
max}=2.6$. The low-loss situation coexists with a high-loss
situation; although a spontaneous total breakdown of the
operational system into the non-operational high-loss phase does
not occur during the simulation, external influences may well
induce such a transition. The second panel has the same $p^{\rm
max}$ but  $(p_{ij}/p_{i} )^{\rm max}=3$. The low-loss situation
is unstable and spontaneously decays to a high-loss situation via
a bubble-nucleation process.}
\end{figure}

\begin{figure}[p]
{\centering
\epsfig{file=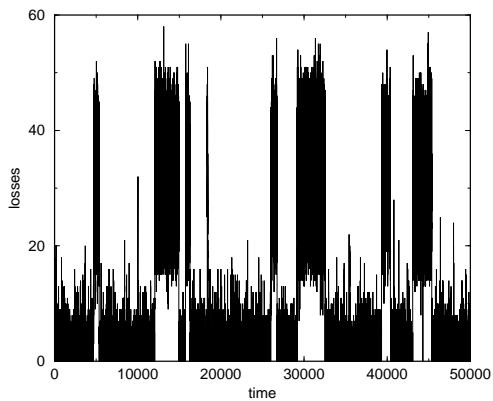,width=6.5cm}
\par}
\caption[]{In a small system ($N=10$), repeated changes between
high- and low-loss situations can be observed.}
\end{figure}

Fig. 3 illustrates the principle of a strain-simulation. In each
case, the system, if in the operational state, is repeatedly put
under external strain by turning off 5 randomly  selected
functioning processes every 1000$^{\rm th}$ time step, and letting
the system evolve under its internal dynamics thereafter. Such a
disturbance can either trigger a breakdown of the system or not.
In the former case, if the system is found fully down 1000 time
steps later, it is reinitialized in the fully operational state
and once more disturbed 1000 steps later.

\begin{figure}[p]
{\centering
\epsfig{file=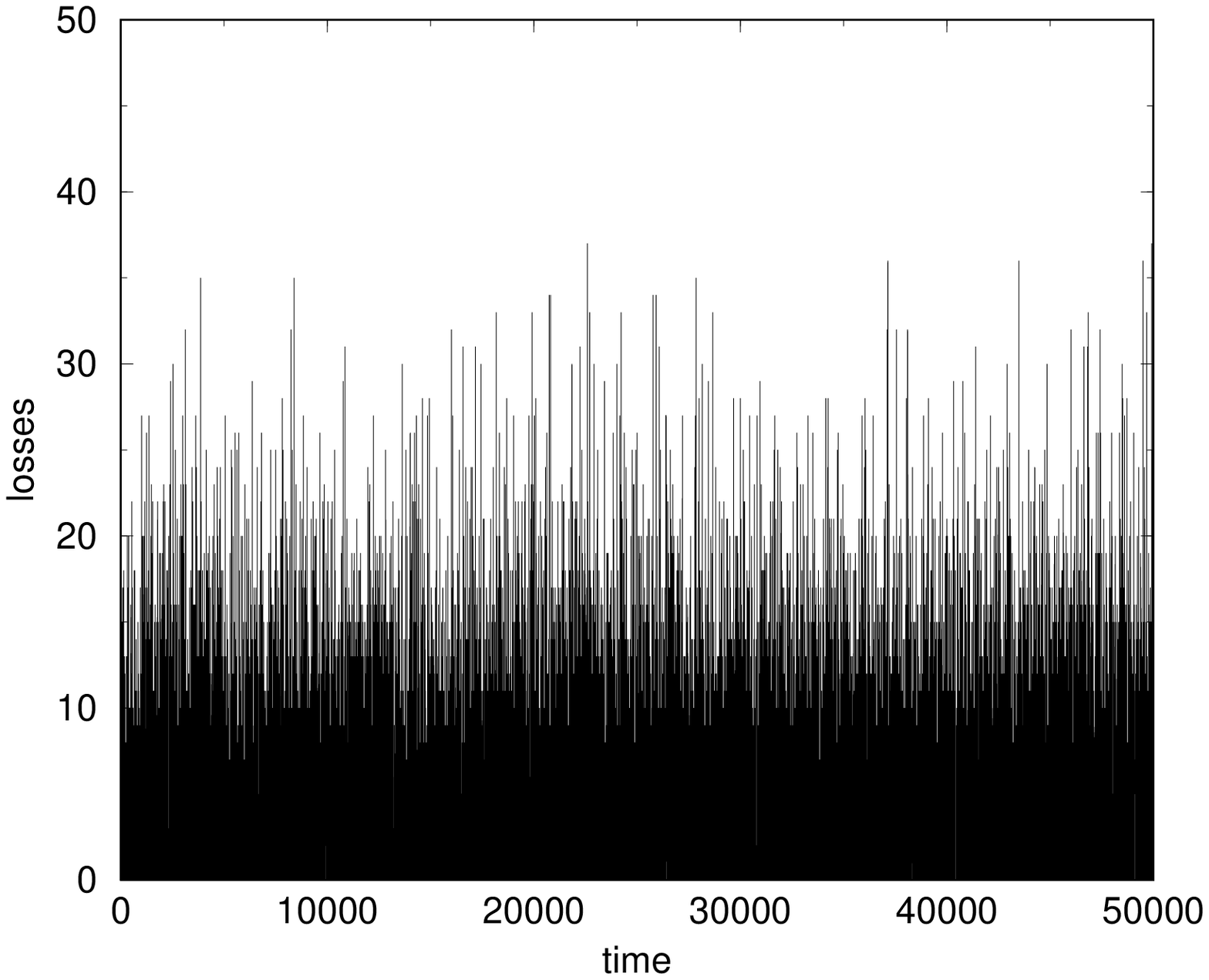,width=6.5cm} \hfil
\epsfig{file=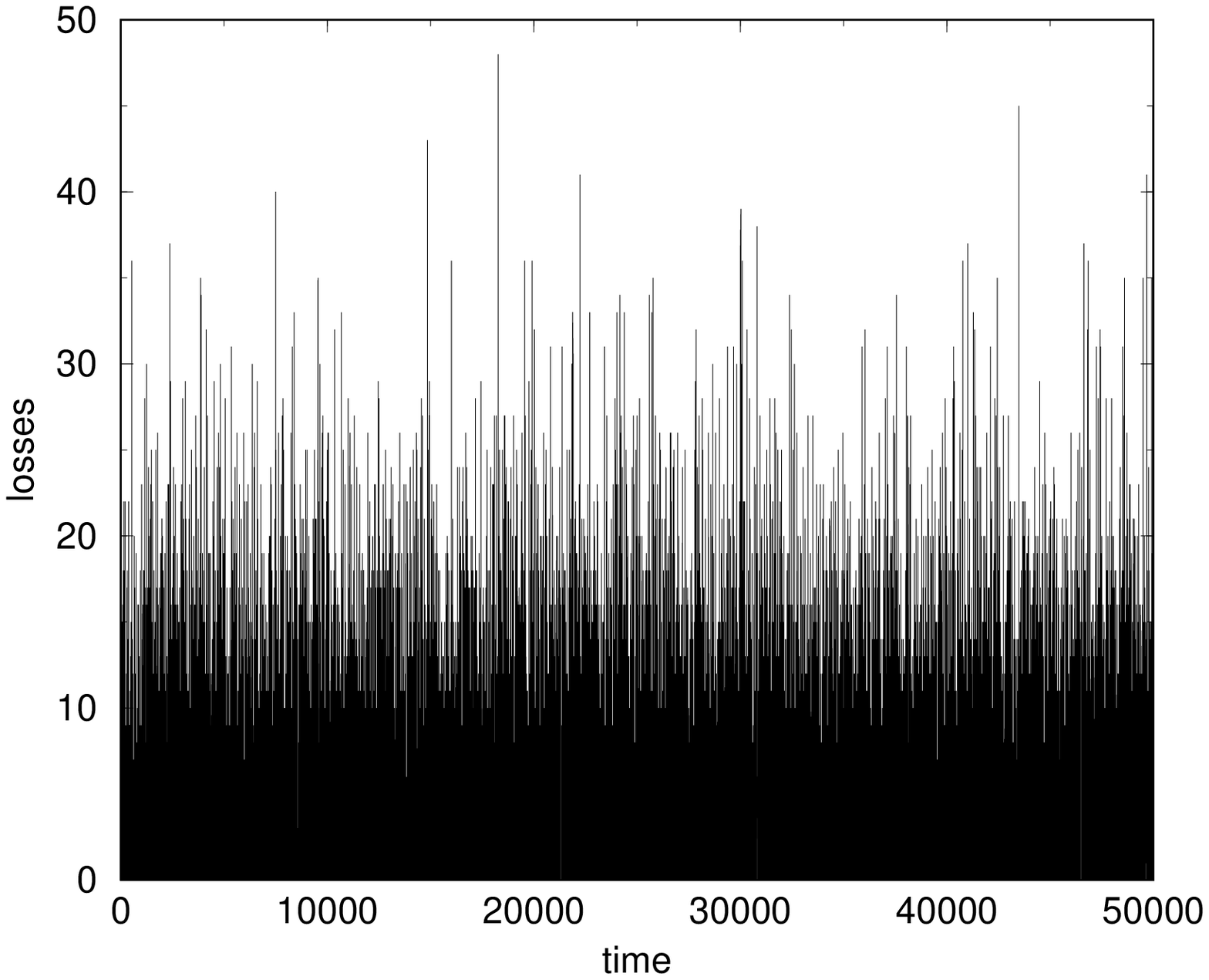,width=6.5cm}
\epsfig{file=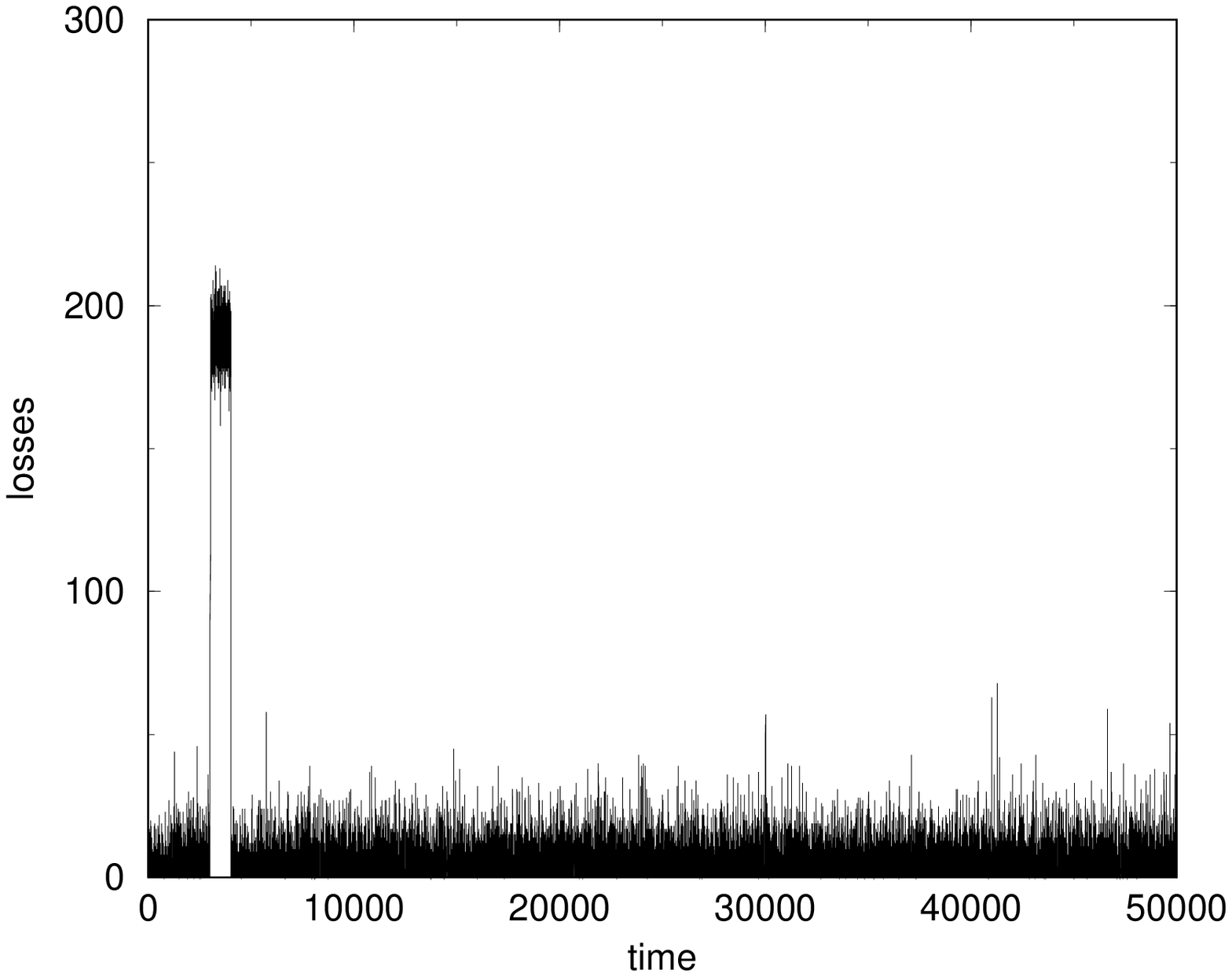,width=6.5cm} \hfil
\epsfig{file=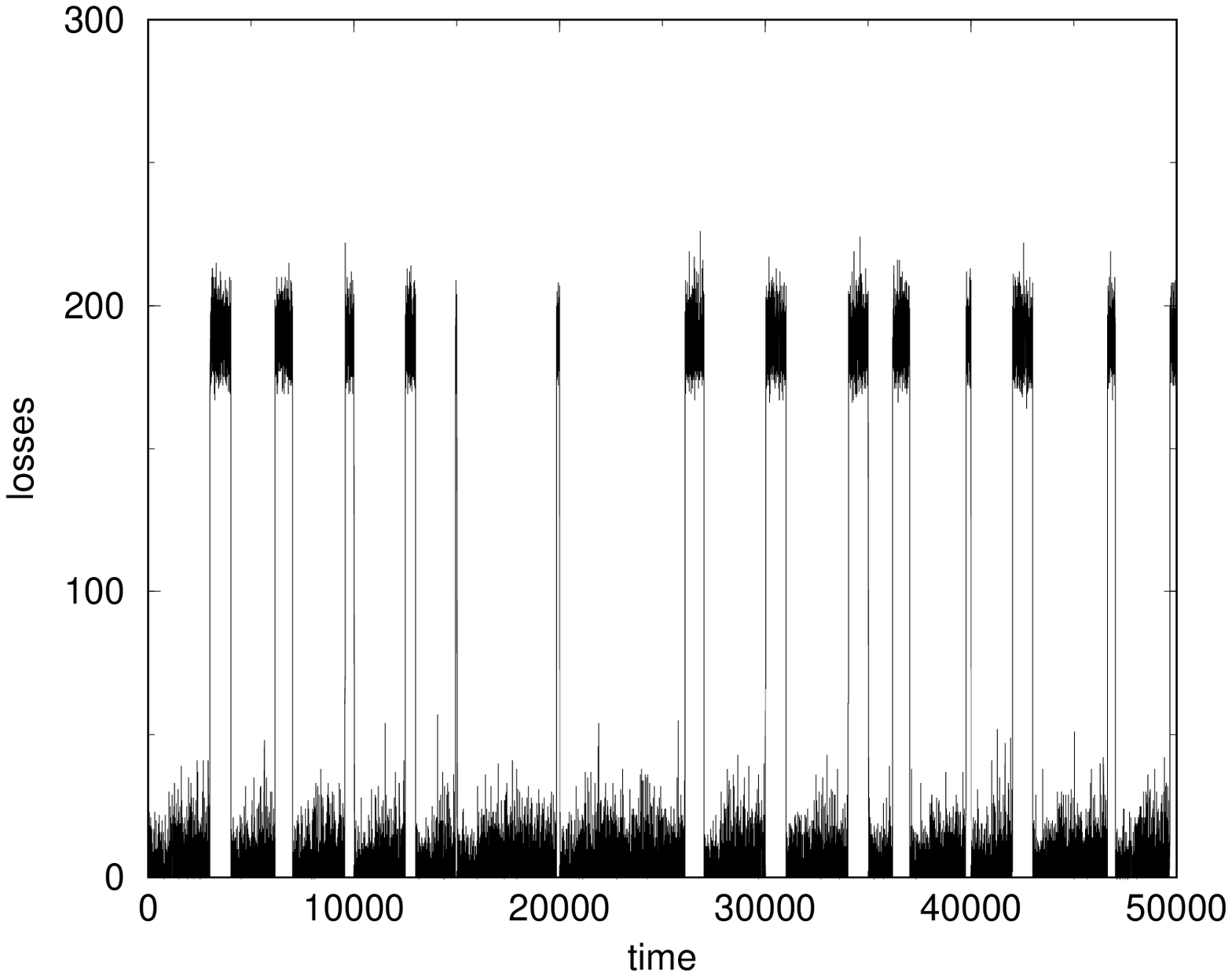,width=6.5cm}
\par}
\caption[]{Strain simulations described in the main text. The
parameters are $p^{\rm max}=0.02$ as in the previous figures and
$(p_{ij}/p_{i})^{\rm max}=2.6$, 2.7, 2.8, and 2.9. }
\end{figure}

\begin{figure}[t]
{\centering
\epsfig{file=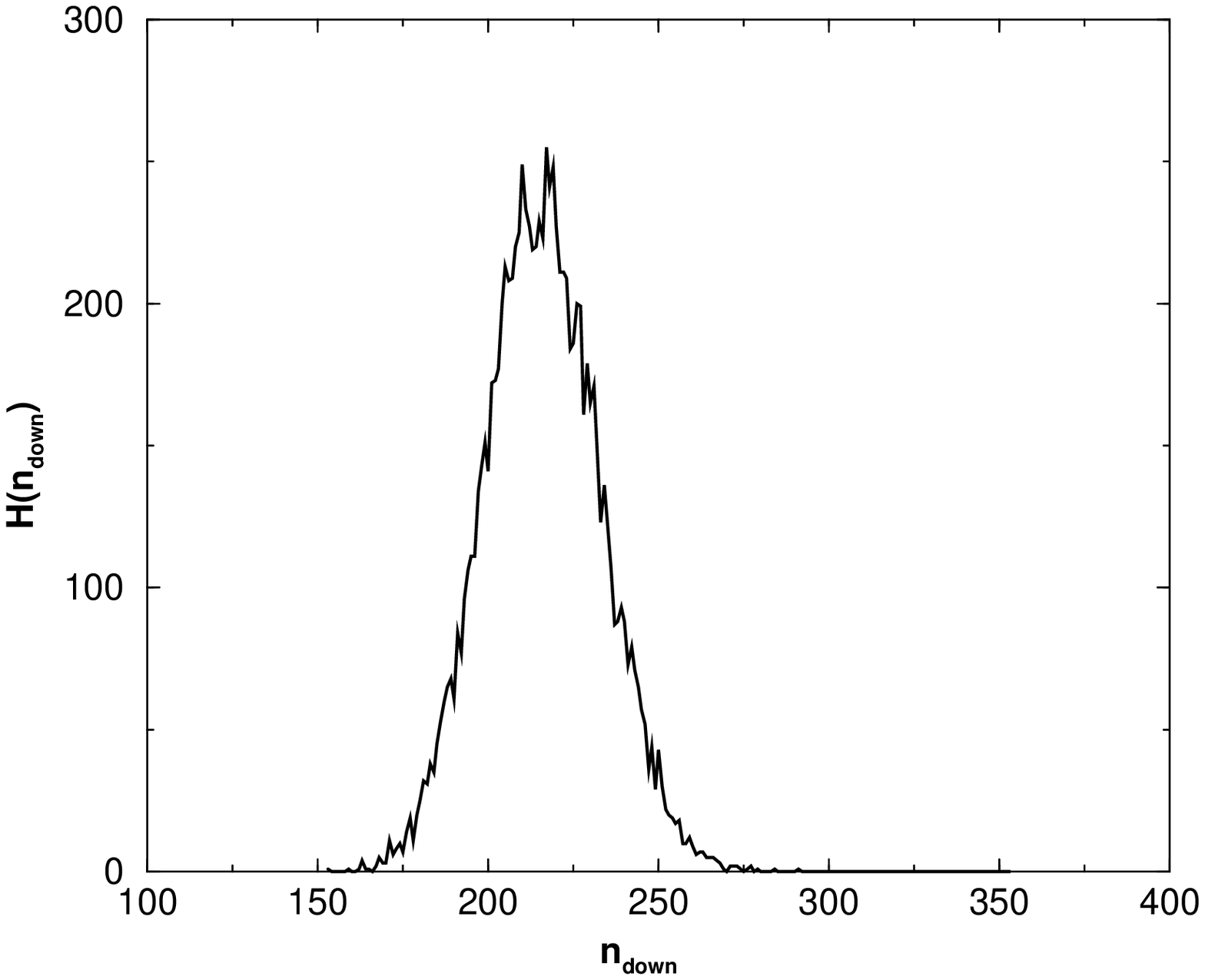,width=6.5cm}\hfil
\epsfig{file=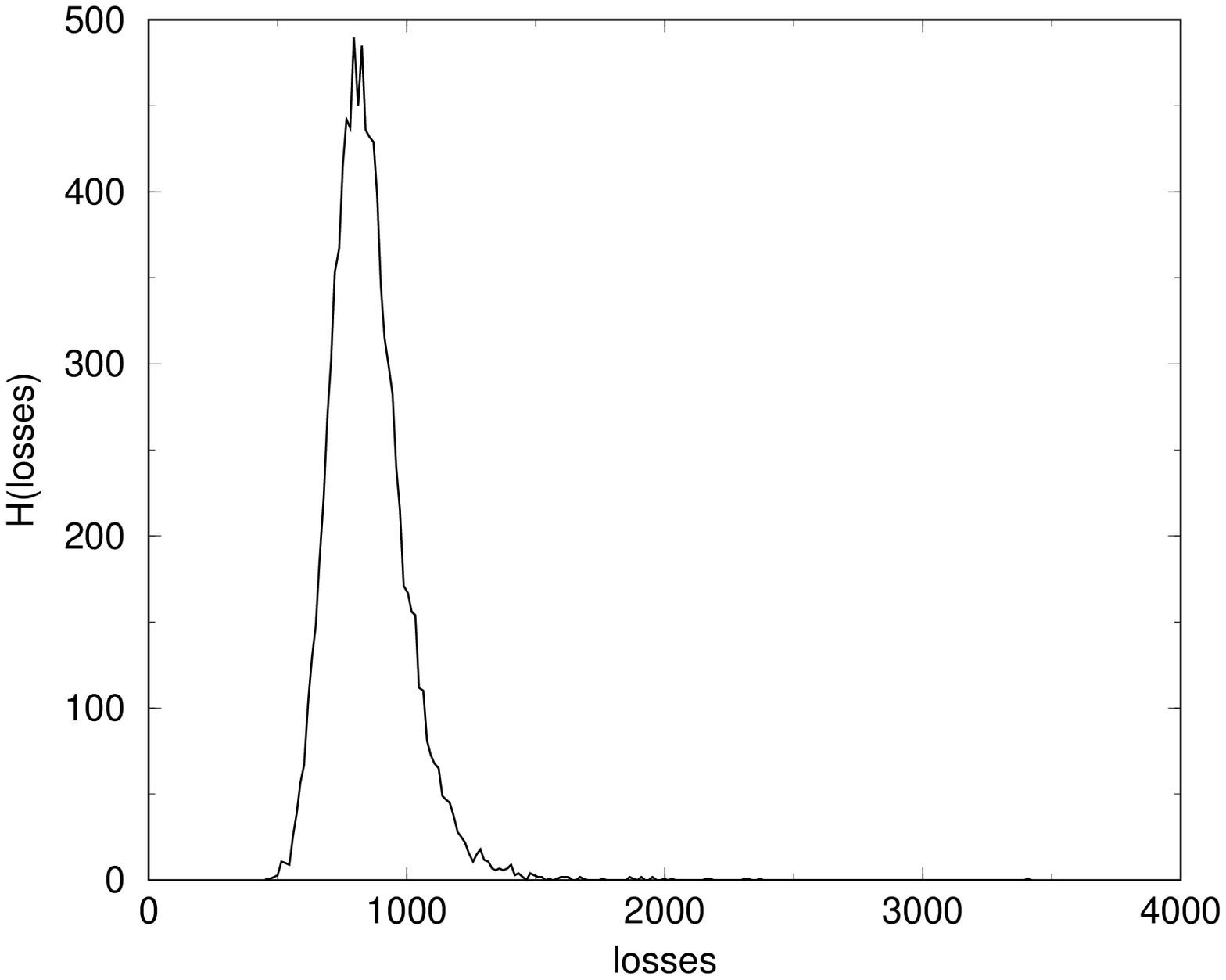,width=6.5cm}
\par}
\caption[]{Frequency distribution of aggregated number of
down-processes during a risk horizon of $T=365 \Delta t$ (first);
loss distribution (second panel). Total time covered was $10^4 T$.
Histograms are not normalized. Parameters of the system are
$N=50$, $p^{\rm max}=0.02$, and $(p_{ij}/p_{i})^{\rm max}=2.5$,
loss severity distributions are taken as log-normal, with
means randomly spread over an interval [0,10] and volatilities 
chosen
randomly as a factor of their respective means, the maximum
factor being 0.4.}
\end{figure}

One observes that the operational low-loss phase which we have
seen in Fig. 1 to coexist with the non-operational phase is
resilient against disturbances of the kind described above. The
same is true if $(p_{ij}/p_{i})^{\rm max}$ is increased to 2.7. At
$(p_{ij}/p_{i})^{\rm max}=2.8$ an external strain succeeds {\em
once\/} during the simulation to trigger breakdown of the net,
whereas at $(p_{ij}/p_{i})^{\rm max}=2.9$ breakdown under
external strain of the given strength is the regular response of
the system (with a few exceptions and occasional spontaneous
recoveries).

Of interest to the risk manager in the end are the total losses
accumulated over a risk horizon, $T$,
\be L(T) = \sum_i L_i(T)\ , \ee
more specifically, the corresponding probability density function.
Fig. 4 presents such distribution of accumulated losses for a
network that remains operational throughout the simulation.  For 
this  simulation we have chosen $T = 365 \Delta t$.  The loss 
distribution 
has an extended tail (barely visible on the scale 
of the data,
with a 99.5\% quantile at 1400, and the largest 
aggregated loss
observed during the simulation over a time span 
of $T$ at 3400, i.e. by
 a factor of more than 3 larger than the 
expected loss for the
chosen risk horizon $T$.

\subsection{Mean-field Solution}

The principle dynamic properties of the model exist independently
of its heterogeneous nature. To exhibit systematic relations
between dynamic features and model parameters, we shall elucidate
them in the simplified setting of a homogeneous network of
identical processes, which we analyze within a mean-field
approximation.

For this we assume a homogeneous coupling, $w_{ij}\longrightarrow
w_0/z$ and $p_{ij} \longrightarrow p_w$, where $z$ is the
coordination number of the graph (taken to be identical at each
vertex), and replace time--dependent quantities in Eq.
(\ref{pdown}) by long-time stationary averages, $n_j(t)
\longrightarrow \langle n_j(t)\rangle$. In a homogeneous system,
averages are independent of the process index $i$ and time, such
that $p_i  \longrightarrow p$ and $\langle n_j(t)\rangle
\longrightarrow n$. This gives the mean-field equation (without
correlation to the common factor, $\rho\equiv 0$)
\bea \label{meanfield}    n = \Phi\left(\Phi^{-1}(p) + w_{0}\, n
\right)\ .\eea

Depending on $p$ and the average coupling strength, $w_0$, this
equation has either one unique solution or three solutions, with
one unstable solution at intermediate $n$, and two stable
solutions, $ n \approx$ 0 or 1. Figure 5 shows stable and
unstable solutions as functions of $w_0$ and $p_{w}$ for a given
value of $p$.

The phase diagram (Fig. 6) summarizes regions in the $p$--$w_0$
plane, where operational and non-operational phases of the
network coexist showing limits of stability of the low-loss and
high-loss solution. For $p$ exceeding a critical value $p_c\simeq
0.0218$ there is a unique disordered phase with relatively large
values of $n$.

\begin{figure}[t]
{\centering
\epsfig{file=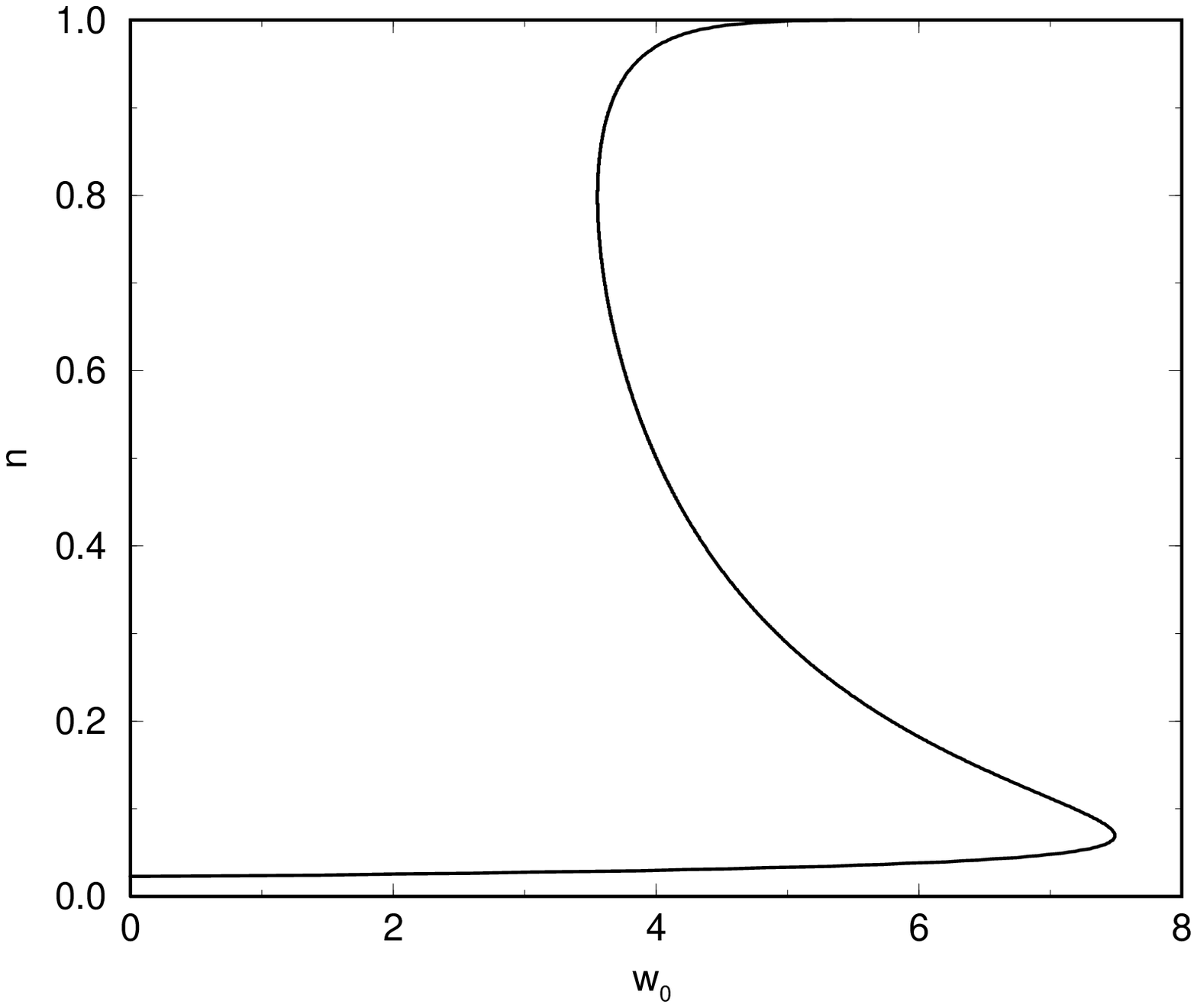,width=6.5cm} \hfil
\epsfig{file=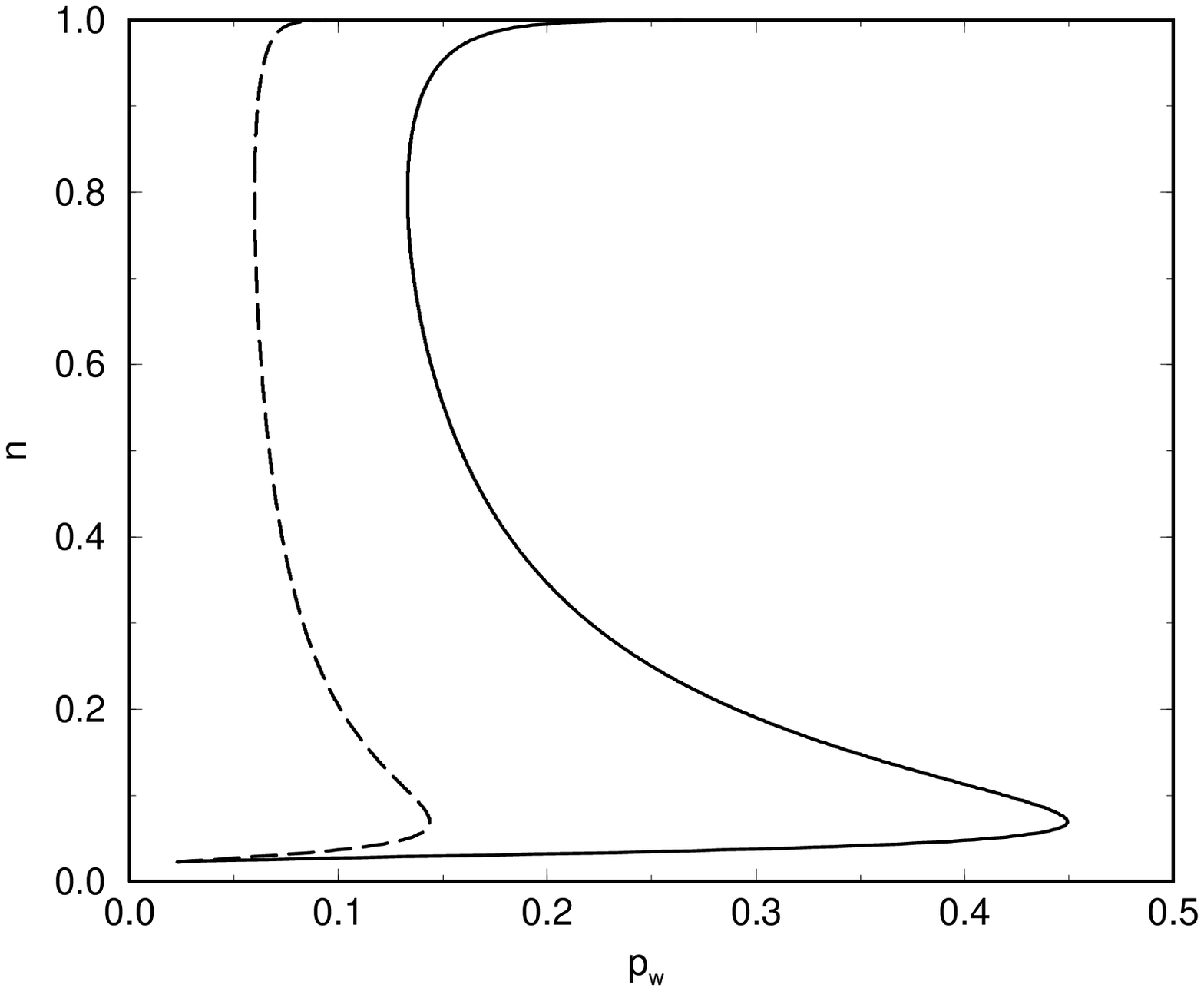,width=6.5cm}
\par}
\caption[]{Mean-field solution for the average fraction $n$ of
down-processes for $p\simeq 0.02$ as a function of the coupling
parameter. The back-turning part is the unstable solution
$n_u(w_0)$. For $n>n_u$ the system is driven towards the
non-operational high-$n$ solution, as long as  $n<n_u$, the
system is driven back to the operational low-$n$ solution. In the
second panel $w_0$ is translated into a conditional probability
$p_w$ under the assumption that the coordination is $z=4$ (right
curve) and $z=8$ (left curve).} \end{figure}

\begin{figure}[h]
{\centering
\epsfig{file=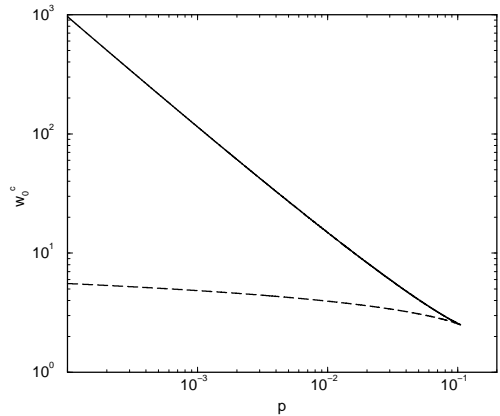,width=6.5cm}
\par}
\caption[]{Mean-field phase diagram for the homogeneous
interacting processes system. At low $p$ operational and
non-operational process networks coexist, separated by a
discontinuous phase transition. Shown are couplings $w_0^c$
corresponding to spinodals which mark instabilities of fully
operational low-loss and non-operational high-loss situations
(upper and lower curves, respectively). Note that hysteresis
effects are implied. The spinodals merge in a critical endpoint,
where the transition is second order.}
\end{figure}

\begin{figure}[h]
{\centering
\epsfig{file=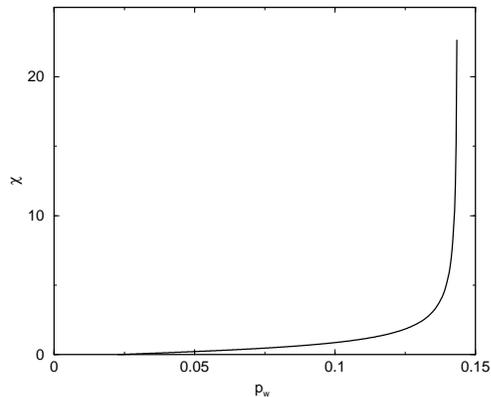,width=6.5cm}
\par}
\caption[]{Susceptibility of the mean-field solution for $p
\simeq 0.02$ as a function of $p_w$ for the homogeneous
interacting processes system at $z=8$. The susceptibility
diverges as the spinodal is approached.}
\end{figure}

For a sufficiently small unconditional failure probability $p$ an
initially running process network will for weak functional
dependence remain in the running state, despite of spontaneous
individual process failures. Such a network is in a functioning
state.

For stronger functional correlation the functional state of the
network becomes unstable. Fluctuations in the number of processes
that are down at any given time can trigger a burst or avalanche
of failures --- a collective phenomenon corresponding to droplet
formation associated with a first order phase transition.

For intermediate degrees of functional dependence, the network
allows for two (meta)stable states, an operational and a
non-operational network. While spontaneous fluctuations in the
number of failing processes will in large networks fail to
trigger transitions between the two states, external strain may
well do, as demonstrated in the strain simulations above. Seen
from the functioning side, the closer the parameters are to an
instability line, the smaller the strain needed to trigger
avalanches of failures leading to the fully defunct state of the
system. This behavior is quantified by the susceptibility $\chi$
(Fig. 7). With support for each process decreasing by an amount
$\delta h < 0$, the fraction of down processes will change by
 \be
\delta n \simeq \chi(w_0,p)\, \delta h \ee
with
\be
\chi(w_0,p) = - \frac{1}{1 - w_0 \Phi'(\Phi^{-1}(p) +w_0 n))}\ ,
\ee
where $\Phi'(x)= \frac{1}{\sqrt{2\pi}}\exp[-x^2/2]$. Note that the
susceptibility is proportional to the sensitivity of the fraction of
down-processes with respect to the unconditional probability of 
process failure.

Triggering complete failures becomes easier close to instability
lines for two reasons: (i) the susceptibility diverges as
instability lines are approached; (ii)  the unstable solution
$n_u$ is closer to the stable equilibrium solution; external
strain only needs to push the system beyond $n_u$ to destabilize
the functioning state.

Such a regime could therefore be dangerous for a network of mutually
supporting processes in a bank. Due to the long periods in one of the
metastable states, the bank would not necessarily realize the potential of
big losses due to bursts and avalanches of process failures. With a basically
unchanged process setup the network could collapse and cause significant
losses, either due to external strain or rare fluctuations of internal
dynamics. Owing to the stability of the metastable states, the bank will then
have to spend a lot of efforts in order to bring the network back to a
functional state, which will cause additional costs.


\section{Conclusion}\label{Sect_Conclude}

In this paper we have outlined how ideas from physics of
collective phenomena and phase transitions can naturally be
applied to model building for operational risk in financial
institutions. Our main point was that functional correlations
between mutually supportive processes give rise to non-trivial
temporal correlation, which could eventually lead to the 
collective
occurrence of risk event in form of burst, avalanches 
and
crashes. For risks associated to process failure (operational
risks) a functional dependence seems to be the appropriate way
for modeling sequential correlations.

From the physics point of view, the appropriate model is
rather simple, being a heterogeneous variant of the well
studied lattice gas model. Despite the heterogeneities, it has
a first order phase transition (driven by average interaction
strength) at sufficiently low average a-priori probability of
process failures, showing coexistence between an overall
functioning state (gas) and a state of catastrophic breakdown
(liquid). As the a-priori probability of process failures
is increased the first order transition ends at a (liquid/gas)
critical point.

One of the most critical lessons for Risk Control from our
analysis is the possible metastability of networks of interacting
processes: The bank would not necessarily realize the potential
of big losses due to bursts and avalanches of process failures,
as there are no detectable precursors to such transitions. With a
basically unchanged process setup the network could collapse and
cause significant losses, either due to external strain or rare
fluctuations of internal dynamics. Owing to the stability of the
metastable states, the bank will then have to spend a lot of
efforts in order to bring the network back to a functional state,
which will cause additional costs. To assess the metastability
banks have to perform stress tests.

It should be noted that realistically the system dynamics after
an overall break-down of a process network would no longer be
the spontaneous internal network dynamics: recovery efforts would
be started, increasing support for each process by a sufficient
amount such as to reinitialize the network in working order.

In a forthcoming publication we will also show that the random
walk model for financial time series commonly used in banks can
naturally be
extended to incorporate functional dependencies
leading to collective effects. This will lead to models which,
while bearing some resemblance with agent-based models of markets
(see, e.g., \cite{agents}) are different from them in other
respects \cite{tobepub}.

{\bf Acknowledgments} We would like to thank U. Anders and
G. J. van~den Brink of Dresdner Bank for valuable discussions and
practical input.



\begin{thebibliography}{99}

\bibitem{RM}
J.P. Morgan~Global Research, {\em RiskMetrics$^{\rm TM}$}, Technical
Document, 4th Edition, URL: {\tt http://www.riskmetrics.com}, (New York, 1996).

\bibitem{BISIIa}
Basel~Committee on~Banking~Supervision, {\em The New Basel
Accord}, Consultive Document, URL: {\tt http://www.bis.org},
(Basel, January 2001).

\bibitem{BISIIb}
Basel~Committee on~Banking~Supervision, {\em Operational Risk},
Consultive Document, URL: {\tt http://www.bis.org}, (Basel,
September 2001).

\bibitem{BISIIc}
Basel~Committee on~Banking~Supervision, {\em The Internal
Ratings-Based Approach}, Consultive Document, URL: {\tt
http://www.bis.org}, (Basel, January 2001).

\bibitem{Frachot01}
P.~Georges A.~Frachot and T.~Roncalli.
{\em The loss distribution approach for operational risk}, Working Paper,
Credit Lyonnais (2000)

\bibitem{Vandenbrink}
G.-J. van~den Brink, {\em Operational Risk, the new challenge for
banks}, Palgrave, (Hampshire (UK), 2001); {\em Die Bedeutung
operativer Risiken f\"ur Eigenkapitalunterlegung und
Risikomanagement}, Tagungsbericht vom Duisburger Bank-Symposium,
Rolfes (Editor) (Duisburg, 2002).

\bibitem{Bouch0}
J.-Ph. Bouchaud, {\em  Power-laws in economy and finance: some ideas from
physics},  e-print, cond-mat/0008103.

\bibitem{Bouch1}
J.-Ph. Bouchaud, 
Physica A {\bf 285}, 18 (2000).

\bibitem{Bouch2}
J.-Ph. Bouchaud and M.~Potters, {\em Theory of Financial Risks},
Cambridge Univ. Press, (Cambridge, 2000).

\bibitem{Sornette1}
A.~Johansen D.~Sornette and J.-Ph. Bouchaud, J. Phys. I France, {\bf 6} 167,
(1996).

\bibitem{Stanly1}
L.A. Amaral M.~Meyer V.~Plerou, P.~Gopikrishnan and H.E. Stanley,
Phys. Rev. E {\bf 60}, 6519 (1999).

\bibitem{Markowitz52}
H.~Markowitz,  {\em Portfolio Selection}, vol.~7. (1952).

\bibitem{Markowitz59}
H.~Markowitz, {\em Portfolio Selection: Efficient Diversification of
Investments}, John Wiley, (New York, 1959).


\bibitem{Nature01}
M.~Scheffer, S.~Carpenter, J.~A. Foley, C.~Folke, and B.~Walker,
Nature, {\bf 413} 591 (2001).

\bibitem{agents}
D. Challet and Y.C. Zhang Physica A {\bf 246}, 407 (1997);
A. Cavagna, J.P. Garrahan, I. Giardina and D. Sherrington, Phys. Rev. Lett.
{\bf 83}, 4429 (1999);
A.C.C. Coolen and J.A.F. Heimel, J. Phys. A {\bf 34}, 10783 (1999);
J. P. Bouchaud, ``Power-laws in economy and finance: some ideas from physics",
cond-mat/0008103, and references therein
J.P. Garrahan, E. Moro, D. Sherrington, Quantitative Finance {\bf 1}, 246
(2001);

\bibitem{tobepub}
R. K\"uhn and P. Neu, manuscript in preparation.
\end{thebibliography}
\end{document}